\documentclass[11pt,a4paper]{article}
\usepackage{jcappub}
\usepackage{bm}
\usepackage{amsmath}
\def\dd{\mathrm{d}}

\def\mcP{\mathcal{P}}

\def\Mpl{M_{\rm Pl}}
\def\GeV{{\rm GeV}}

\def\mseff{m_{\sigma,{\rm eff}}}
\def\ms{m_{\sigma}}

\def\fnl{f_{\rm NL}}
\def\taunl{\tau_{\rm NL}}
\def\hatr{\hat{r}}
\def\etas{\eta_\sigma}

\title{
Curvaton in large field inflation
}
\author[a,b]{Tomohiro Fujita}
\author[c]{Masahiro Kawasaki}
\author[c]{Shuichiro Yokoyama}
\affiliation[a]{Kavli Institute for the Physics and Mathematics of the
Universe (Kavli IPMU), TODIAS,  the University of Tokyo, 5-1-5
Kashiwanoha, Kashiwa, 277-8583, Japan}
\affiliation[b]{Department of Physics, University of Tokyo, Bunkyo-ku
113-0033, Japan}
\affiliation[c]{Institute for Cosmic Ray Research, University of Tokyo,
5-1-5 Kashiwa-no-Ha, Kashiwa, Chiba, 277-8582, Japan}
\emailAdd{tomohiro.fujita@ipmu.jp}
\emailAdd{kawasaki@icrr.u-tokyo.ac.jp}
\emailAdd{shu@icrr.u-tokyo.ac.jp}

\abstract{
We comprehensively explore the quadratic curvaton models
in the chaotic inflation.
In the light of the BICEP2 result $r\approx 0.2$,
all model parameters and relevant observables
are computed.
It is found the curvaton field value is constrained into a
narrow range, $\sigma_* = \mathcal{O}(10^{-2}$-$10^{-1})$
and the running of the spectral index is $n_s' \gtrsim -10^{-3}$.
We show that if the curvaton is added, the models are heavily degenerated on the $n_s$ - $r$ plane.
However, introducing a new plane, the degeneracy can be resolved.
To distinguish the curvaton models, precise measurements of not only
$r$ but also $n_s'$ and the tensor tilt $n_T$ are required.
}

\arxivnumber{1404.0951}

\begin{document}

\begin{flushright}
ICRR-Report-676-2014-2
\\
IPMU 14-0089
\end{flushright}

\maketitle

%
%
%
\section{Introduction}

The inflation paradigm is accepted as an increasingly plausible scenario
of the primordial universe. Although the integrated theory that describes high energy physics including inflation is not yet identified,
the candidates (e.g. the supergravity theory) predict the existence of many scalar fields and one of them can play a role of an inflaton.
Since a light scalar field acquires its fluctuation during inflation,
it is natural to expect that not only the inflaton but also the other
scalar fields contribute to the generation of  primordial cosmic
perturbations. Such scenarios where the contribution of additional scalar fields to the observed perturbations are significant are known as 
``the curvaton-like models".
In curvaton models, an additional scalar field (curvaton) has only a tiny fraction of the total energy density during inflation 
but its energy fraction increases after inflation and 
produces the curvature perturbation when it decays~\cite{Enqvist:2001zp,Lyth:2001nq,Moroi:2002rd}.

Recently, BICEP2 experiment has reported a large tensor-to-scalar ratio,
$r\approx 0.2$~\cite{Ade:2014xna}.
\footnote{There are related papers about curvaton scenario
in the light of BICEP2~\cite{Byrnes:2014xua,Sloth:2014sga}.}
If it is confirmed, it indicates that 
the Hubble parameter during inflation is $\mathcal{O}(10^{14})$GeV and the large field inflation models are strongly favored. 
With such a large Hubble parameter, a curvaton acquires a large fluctuation.
However, an inflaton in the large field models also significantly contributes to the perturbations and their values can be explicitly computed if the inflation model is fixed.
Because these perturbations generated by the two sources should be compatible with the observations, we can put strong constraints on the curvaton models.

Moreover, considering that the curvaton contribution may be relevant,
it is important to find a way to determine the curvaton model
in addition to the inflation model.
As for the discrimination of the inflation models, 
the spectral index $n_s$ and the tensor-to-scalar ratio $r$ are useful
and the $n_s$- $r$ plane is frequently utilized.
Once a curvaton is added, however, a wide region on the the $n_s$- $r$ plane can be reproduced by various models and the plane is not sufficient to
distinguish the curvaton models. Therefore, when curvaton models are discussed,
it is necessary to investigate more observables and the verifiability of the models.

In this paper, we comprehensively explore the two curvaton models in the chaotic inflation~\cite{Linde:1983gd, Linde:2005ht}. 
In one model, the curvaton has only its constant mass, $\ms$. 
In the other model, the curvaton has the Hubble induced mass as well as its intrinsic mass. We investigate not only the curvaton mass and $n_s$ 
but also the running $n_s'$, the curvaton field value during inflation $\sigma_*$,
the non-linear parameters $\fnl, \taunl$, its energy fraction at its decay
$\hatr$ and its decay rate $\Gamma_\sigma$ in the two models.

We show these curvaton models can realize a wide range of $n_s$ and $r$ values. Furthermore, these models are heavily degenerated on the $n_s$-$r$ plane. Then we discuss the running of the scalar spectrum $n_s'$ and 
a combination of $r$ and the tilt of the tensor spectrum $n_T$ to
discriminate the curvaton models.
It is demonstrated that these curvaton models
are distinct on the $n_s'$ - $(n_T/r)$ plane.
Thus we found the new plane is useful to discuss the difference between 
curvaton-like models in the same way as the $n_s$-$r$ plane for inflation models.
It is also found that the field value of the curvaton 
during inflation is constrained into $\mathcal{O}(10^{-2}$-$10^{-1})\Mpl$.

The rest of this paper is organized as follows.
In section~\ref{review}, first we briefly review the chaotic inflation model,
and then we show the implication of the curvaton scenario
with the relatively large tensor-to-scalar ratio.
We also mention the degeneracy of the models in 2-dimensional parameter space
of the tensor-to-scalar ratio and the spectral index of the curvature perturbation.
In section \ref{distinguish}, we explore the possibility to resolve the degeneracy
by introducing a new 2-dimensional parameter space.
In section \ref{NG}, the non-linear parameters and a constraint on the curvaton
field value are calculated. Section \ref{Conclusion} is devoted to the conclusion.

\section{Implication for the curvaton scenario with the chaotic inflation}
\label{review}
\subsection{Brief Review on the chaotic inflation}

In this section, we briefly review the chaotic inflation model.
The inflaton potential is given by~\cite{Linde:1983gd,Linde:2005ht}
\begin{equation}
U(\phi) = \frac{1}{p} \frac{\phi^p}{M^{p-4}}. 
\label{inflaton potential}
\end{equation}
This potential can be naturally realized in the supergravity setup~\cite{Kawasaki:2000yn, Kawasaki:2000ws,Kallosh:2010ug, Kallosh:2010xz}. 
For such kind of potential, slow-roll parameters are given by
\begin{eqnarray}
\epsilon &\equiv& {\Mpl^2 \over 2} \left( {U' \over U} \right)^2 = {p^2 \Mpl \over 2 \phi^2}, \cr\cr
\eta &\equiv & \Mpl^2 {U'' \over U} = p (p - 1) {\Mpl^2 \over \phi^2} = {2(p - 1) \over p} \epsilon ,
\label{slowroll}
\end{eqnarray}
where a prime denotes the derivative in terms of $\phi$.
In the slow-roll approximation where $\epsilon, \eta \ll 1$, the e-folding number during inflation is evaluated as
\begin{equation}
N \simeq  \Mpl^{-2}\int^\phi_{\phi_{\rm end}}
\frac{U}{U'}d\phi
\simeq \frac{1}{2p\Mpl^2} \left(\phi^2-\phi^2_{\rm end} \right)
=\frac{\phi^2}{2p\Mpl^2}-\frac{p}{4},
\end{equation}
where we define the end of inflation as $\epsilon_{\rm end} = 1$.
Then the slow-roll parameters are written in terms of e-folding number as
\begin{align}
\epsilon  =\frac{p}{4N+p}.
\label{eps}
\end{align}
%
%
%
The curvature perturbation 
generated by the inflaton fluctuation is simply given by
\begin{align}
\mcP_\zeta^{(\phi)}(k) = \left(\frac{H_*^2}{2\pi \dot{\phi}_* }\right)^2
=\frac{H_*^2}{ 8 \pi^2 \epsilon \Mpl^2},
\label{inflaton P_zeta}
\end{align}
where a dot denotes the derivative in terms of cosmic time and
the subscript ``$*$" denotes quantities at the horizon crossing of  the mode
with a corresponding wave number $k$ while the superscript ``$(\phi)$" denotes quantities generated by the inflaton.
By using the slow-roll parameter (or e-folding number), the spectral index $n_s$ 
is also predicted as
\begin{align}
n_s^{(\phi)}-1 &\equiv \frac{\dd \ln \mcP_\zeta^{(\phi)}}{\dd \ln k}
= -2 \left( \frac{2+p}{p} \right) \epsilon = -2 \left( {2 + p \over 4N + p}\right)
<0,
\label{inf run}
\end{align}

Gravitational waves are also produced during inflation
and its power spectrum is predicted as
\begin{equation}
\mcP_g (k) = \frac{8}{\Mpl^2} \left(\frac{H_*}{2\pi} \right)^2.
\label{P_g}
\end{equation}
%
The ratio between $\mcP_\zeta^{(\phi)}$ and $\mcP_g$
can be simply related with the slow-roll parameter as
\begin{equation}
r^{(\phi)} \equiv \frac{\mcP_g}{\mcP_\zeta^{(\phi)}} =16\epsilon.
\label{inflaton r}
\end{equation}
%
%
%
%
%

\subsection{Curvaton contribution to the tensor-to-scalar ratio and the spectral index}
\label{scalar perturbations}

In this section, we generally consider an additional contribution
to the scalar perturbation in curvaton scenario.
However, we note that the results of this section are applicable to not only curvaton models
but also the other type of the curvaton-like models (e.g. the modulated reheating models~\cite{Dvali:2003em, Kofman:2003nx, Zaldarriaga:2003my}).

%
%

\begin{figure}[tbp]
\begin{center}
  \includegraphics[width=100mm]{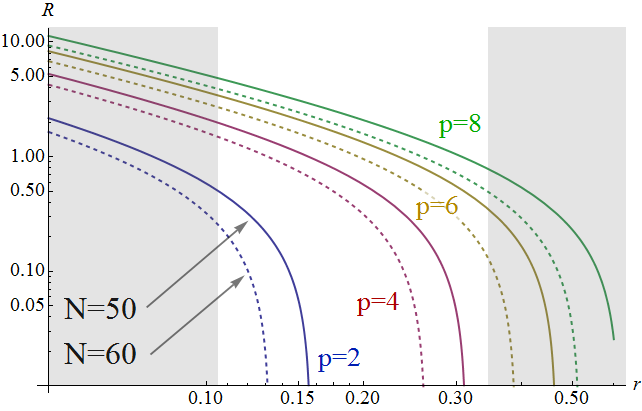}
\end{center}
 \caption
 {The ratio between the contribution to the curvature perturbation of the inflaton and that of another source,
 $R\equiv \mcP_\zeta^{(\sigma)}/\mcP_\zeta^{(\phi)}$. 
 From bottom to top, the lines denote the $\phi^p$ models with $p=2, 4, 6$ and $8$.
 Solid lines and dashed lines represent the cases of $N=50$ and $N=60$, respectively.
Grey shaded regions are excluded by BICEP2 experiment at the $2\sigma$ level.}
 \label{R plot}
\end{figure}

First, according to the $\delta N$ formalism~\cite{Sasaki:1995aw,Wands:2000dp,Lyth:2004gb}, the power spectrum of the curvature perturbation which is sourced not only from the inflaton fluctuation
but also the fluctuation of the additional scalar filed such as curvaton, $\sigma$, 
is given by~\cite{Ichikawa:2008iq, Enqvist:2013paa}
\begin{equation}
\mcP_\zeta (k) = (N_\phi^2+N_\sigma^2)\left(\frac{H_*}{2\pi}\right)^2
\equiv
(1+R)\mcP^{(\phi)}_\zeta,
\label{P_zeta}
\end{equation}
where $N_a \equiv \partial N / \partial a  (a=\phi, \sigma) $ and we define $R$ as the ratio between $\mcP_\zeta^{(\phi)}$ and $\mcP_\zeta^{(\sigma)}$,
\begin{equation}
R \equiv \frac{\mcP_\zeta^{(\sigma)}}{\mcP_\zeta^{(\phi)}} =\left( \frac{N_\sigma}{N_\phi}\right)^2. 
\end{equation}
Note that the superscript ``$(\sigma)$" denotes the contribution of the curvaton
in the following sections while it can be replaced by the other sources.
%
%
%
From the current cosmic microwave background (CMB) observations,
we know the observed value of the amplitude of the scalar curvature perturbation
as $\mcP_\zeta = 2.2\times 10^{-9}$~\cite{Ade:2013uln, Ade:2013ydc}.
On the other hand, as we have shown, the amplitude of the gravitational waves produced during inflation
can be written only in terms of the inflationary Hubble parameter, $H_*$, as eq. (\ref{P_g}) and hence
the tensor-to-scalar ratio given by $r \equiv \mcP_g / \mcP_\zeta$ is directly related with $H_*$ as 
\begin{equation}
H_* = 1.1\times 10^{14} \GeV \sqrt{\frac{r}{0.2}}.
\label{H value}
\end{equation}
The tensor to scalar ratio in case with  the additional contribution to the scalar perturbation due to
the curvaton fluctuation is predicted as 
\begin{equation}
r =  \frac{r^{(\phi)} }{1+R} =  \frac{16\epsilon}{1+R}
\quad \Longleftrightarrow \quad
R = \frac{16}{r} \left( {p \over 4N + p} \right)-1.
\label{r}
\end{equation}
Therefore the fraction of the curvaton contribution, $R$, is fixed by $r, p$ and $N$. In fig.~\ref{R plot}, we plot $R$ as a function of $r$ for each $p$ and $N$.
One can see that $R$ has a finite value in a limited range of $r$ depending on $p$.
This is because  $r$ always decreases from $r^{(\phi)}$ due to a positive $R$
and $r\ge r^{(\phi)}$ cannot be realized
\footnote{See, however, ref.~\cite{Kobayashi:2013awa} in which
the realization of a negative $R$ by a curvaton is discussed.}.
In other words, with an additional contribution to $\mcP_\zeta$,
$r$ is smaller than the prediction of the single slow-roll case;
\begin{equation}
R>0 \quad \Longrightarrow \quad
r < r^{(\phi)}=16\epsilon.
\end{equation}
Here we list the upper bound on $r$ where the curvaton can work,
\begin{align}
N=50:\qquad
r_{(p=2)}^{(\phi)} &= 0.16,
\quad
r_{(p=4)}^{(\phi)} = 0.31,
\quad
r_{(p=6)}^{(\phi)} = 0.46,
\quad
r_{(p=8)}^{(\phi)} = 0.62,
\label{r50}
\\N=60:\qquad
r_{(p=2)}^{(\phi)} &= 0.13,
\quad
r_{(p=4)}^{(\phi)} = 0.26,
\quad
r_{(p=6)}^{(\phi)} = 0.39,
\quad
r_{(p=8)}^{(\phi)} = 0.52.
\label{r60}
\end{align}

\begin{figure}[tbp]
  \hspace{-2mm}
  \includegraphics[width=75mm]{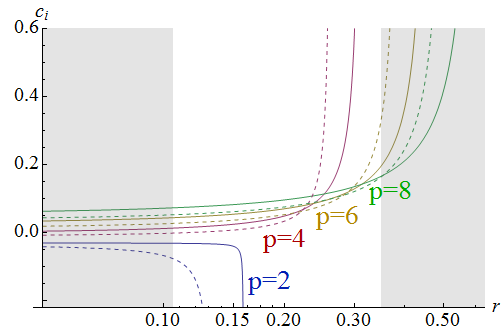}
    \hspace{5mm}
    \includegraphics[width=75mm]{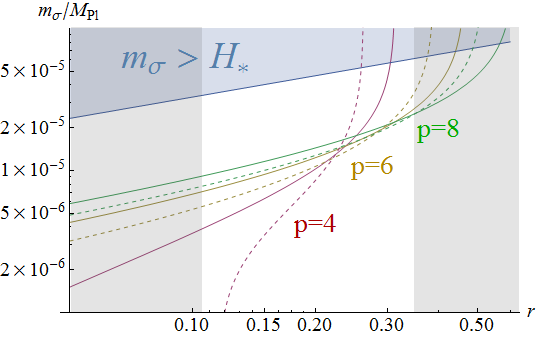}
 \caption
 {The left panel shows the coefficient of the HIM $c_i$
 in the model with $V_H(\sigma)$.
 The right panel shows the intrinsic mass $\ms$ in the model with $V_I(\sigma)$. The spectrum index $n_s-1=-0.04$ is used. 
The e-folding number $N$ is 50 (solid lines) and 60 (dashed lines).
In the blue shaded region, the intrinsic mass of $\sigma$ exceeds $H_*$.
 The $p=2$ line does not appear in the right panel because a positive $m_\sigma^2$
 gives only a bluer spectral index than the single-field case.
 For the same reason, in cases with $p=4, N=60$, the additional contribution to the curvature perturbations in the case with only $\ms^2$ cannot work for $r<0.12$.
}
 \label{ci plot}
\end{figure}

Next, let us focus on the spectral index of the scalar curvature perturbations.
In order to investigate the spectral index,
we need to specify the potential for the curvaton-like field which gives the additional contribution to the curvature perturbations.
Here, we consider the following two models of curvaton,
\begin{eqnarray}
V_H (\sigma ) &= \frac{1}{2}c H^2 \sigma^2,\qquad
V_I (\sigma) &= \frac{1}{2}\ms^2 \sigma^2.
\end{eqnarray}
In the former model, we consider the Hubble induced mass (HIM)~\cite{Matsuda:2007av, Kobayashi:2013bna}.
Note that during inflation, a scalar field generally acquires the HIM through the supergravity effect which is originated from the breaking of supersymmetry due to the inflation energy density~\cite{Stewart:1994ts, Dine:1995uk}.
A parameter which characterizes the amplitude of HIM, $c$, is a constant during inflation,
$c=c_i$, but it takes a different value in 
the successive matter dominant (inflaton oscillation) era and
the radiation dominant era~\cite{Kawasaki:2011zi,Kawasaki:2012qm, Kawasaki:2012rs}
\footnote{
In order for the curvaton to oscillate later, it also has its intrinsic mass even in the HIM model while the intrinsic mass is assumed to be negligible
compared with the HIM during inflation.
It is known that the parameter $c$ during the radiation dominated era is typically $\mathcal{O}(10^{-3}-10^{-2})$~\cite{Kawasaki:2011zi,Kawasaki:2012qm, Kawasaki:2012rs} and hence
not the Hubble induced mass but the intrinsic mass $\ms$ drives the curvaton oscillation, $H_{\rm osc} \simeq \ms$.}.
We assume that both $c_i H^2$ and $\ms^2$ are much smaller than $H^2$ during inflation,
in order for $\sigma$ to acquire quantum fluctuations as
$\mcP_{\delta \sigma}(k) = \left(H_*/2\pi\right)^2$ in both cases.
Then, the spectral index of the curvature perturbation for such case is calculated as~\cite{Enqvist:2013paa}
\begin{align}
n_s-1
&= - \frac{1}{1+R}\frac{2 (2+p)}{4N + p}
+\frac{R}{1+R} \left[-2\epsilon+ \frac{2\mseff^2}{3H_*^2}  \right] ,
\label{tilt}
\end{align}
where $\mseff^2$ denotes $c_i H_*^2$ or $\ms^2$ depending on the models.
It can be easily checked that for $R\to \infty$ the pure curvaton result
and for $R \to 0$ the pure inflaton result
are reproduced. 
It should be noted that if $\mseff^2$ is positive (negative),
it makes the spectral index bluer (redder).  
By solving eq.~\eqref{tilt} with respect to $\mseff^2$ and using eqs. (\ref{eps}) and (\ref{r}), one obtains the curvaton mass as
%
\begin{equation}
\mseff^2 
=\frac{3}{2} \left( n_s - 1 +  \frac{2 p}{4N + p} + \frac{r}{4 p }\right) \left( 1 - \frac{r(4N+p)}{16 p}\right)^{-1} H_*^2.
\label{mseff}
\end{equation}
The r.h.s in eq.~\eqref{mseff} contains only the observables ($n_s, r, H_*$) and the inflation model parameters ($p, N$). Thus the curvaton mass can be determined by the present observation results if the inflation model is fixed.
In fig.~\ref{ci plot}, $c_i$ and $\ms$ are plotted.
%
Although the HIM and the intrinsic mass work in the similar ways, only the former squared can be negative and help the chaotic inflation models which predict larger (bluer) tilts than the observation.
Therefore, in the right panel of fig.~\ref{ci plot}, the $p=2$ case does not appear and the $p=4$ and $N=60$ (dashed line) case cannot has a finite
$\ms$ for $r<0.12$. On the other hand, the $p=2$ lines rapidly drops in the left panel because the additional contribution to the scalar curvature perturbation can work only for $r<r^{(\phi)}$.

\subsection{Degeneracy on the $n_s$-$r$ plane}
\label{degeneray}

It is known well that the $n_s$ - $r$ plane is useful to distinguish single slow-roll
inflation models. Especially, the prediction of the chaotic inflation model can be expressed as a definite point on the plane if the e-folding number $N_*$ is given.
Therefore the constraint on the plane is a powerful way to favor or exclude
the model. However, if the curvaton is added to the chaotic inflation,
since the curvaton mass $\mseff^2$ during inflation and its relative contribution
to the curvature perturbation $R$
are the additional parameters, the model prediction occupies a finite region
on the $n_s$-$r$ plane.
One can eliminate $R$ from eqs.\eqref{r} and \eqref{tilt} and obtains,
\begin{align}
\Delta r &= -\frac{16p \epsilon}{4\epsilon+ 2p\eta_\sigma}\Delta n_s,
\qquad
\left(\Delta r \equiv r-r^{(\phi)}<0,\quad
\Delta n_s \equiv n_s- n_s^{(\phi)},\quad
\eta_\sigma \equiv \frac{\mseff^2}{3H_*^2}\right).
\end{align}
This equation describes how the curvaton can change the prediction
for \{$n_s,r$\} of the original $\phi^p$ chaotic inflation models.
If the curvaton mass is negligible ($\eta_\sigma\to0$), one finds
$\Delta r =-4p\Delta n_s$.
A finite curvaton mass changes the gradient of the line.
For $\eta_\sigma <-2\epsilon/ p$, the gradient becomes positive.
Note since $R$ is positive in our scenario, $\Delta r$ is always negative.
%
\begin{figure}[tbp]
  \begin{center}
  \includegraphics[width=120mm]{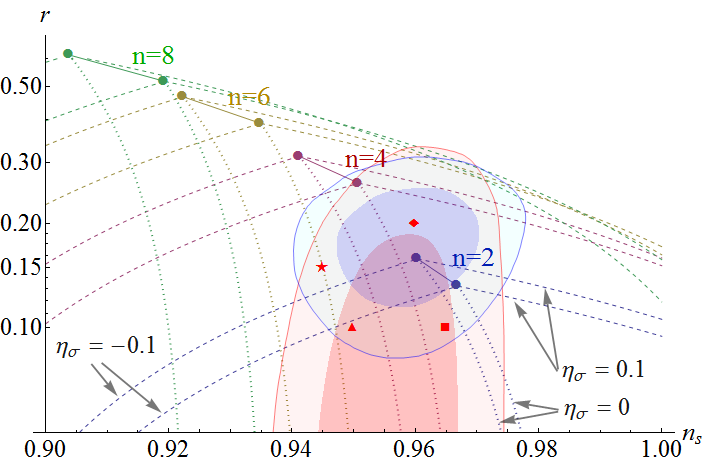}
  \end{center}
   \caption
 {The accessible region of the quadratic curvaton models in the $\phi^p$ chaotic inflation. The pairs of circles represent the prediction of the original chaotic models for $N=50$ (left) $N=60$ (right). The curvaton with appropriate $R>0$ and $|\eta_\sigma|<0.1$ can produce \{$n_s,r$\} in the region below the dashed lines.
Since the dotted lines denote the case of $\etas=0$,
only the curvaton with a negative mass squared generates the region to the left of the dotted lines. Blue and red shaded regions
represent BICEP2 and Planck constraints, respectively~\cite{Ade:2014xna}.
For example, the point of the red star, \{$n_s, r\} =\{0.945,0.15\}$, cannot be realized
by the $p=2$ chaotic + curvaton model, but the $p=4$ model with a negative mass square curvaton and the $p=6, 8$ model with a positive mass curvaton
can reproduce it. One can see that in the observationally favored region,
the models are heavily degenerated.
 }
 \label{help}
\end{figure}
Fig.~\ref{help} shows the region where the curvaton models in the chaotic inflation can access on the \{$n_s,r$\} plane.
Since in the HIM model the curvaton has a negative mass squared
the accessible region is larger than the intrinsic mass model.  
The figure indicates that the models are heavily degenerated
especially in the observationally favored region.

Let us stress that the results of this section are applicable to not only curvaton models
but also the other type of the curvaton-like models.
\section{Distinguish the models?}
\label{distinguish}

In the previous section, we mention that
mixed scenarios, where both of the curvaton and inflation fluctuations
contribute to the curvature perturbation, 
are degenerate
in $n_s$ - $r$ plane known as a useful parameter space to distinguish single slow-roll
inflation models.
Hence, here we investigate a new parameter space where
we would distinguish such mixed scenarios.

\subsection{Tensor tilt and the running of the spectral index of the curvature perturbations}

First, let us consider the tensor tilt $n_T$.
Taking the derivative of eq.~\eqref{P_g}, one finds
\begin{equation}
n_T \equiv \frac{\dd\ln \mcP_g}{\dd\ln k} = -2\epsilon.
\label{nt}
\end{equation}
Combining the above expression for $n_T$ with eq.~\eqref{inflaton r}, we obtain so-called ``the
consistency relation" of single slow-roll inflation models as~\cite{Liddle:1992wi}
\begin{equation}
r^{(\phi)}=-8n_T.
\end{equation}
On the other hand, once the additional contribution to the curvature perturbations exists,
the tensor-to-scalar ratio is modified as shown in the previous section and hence
the above consistency relation would be violated as
\begin{eqnarray}
r = \frac{r^{(\phi)}}{1+R} = \frac{ - 8 n_T}{1+R}.
\end{eqnarray}
In other words, 
through the above expression,
$R$ can be observationally determined
by the combination of the tilt of the tensor spectrum $n_T$
and the tensor-to-scalar ratio $r$ as
\begin{equation}
R=-1-8\frac{n_T}{r}.
\label{intrinsic r}
\end{equation}

On the other hand, the running of the spectral index of the curvature perturbation
in the standard single slow-roll inflation is given by\footnote{When one computes the derivatives of the slow-roll parameters,
$\dd\epsilon/\dd \ln k= 2\epsilon(2\epsilon-\eta)=4\epsilon^2/p$ is useful.}
\begin{align}
n_s'^{(\phi)} &\equiv \frac{\dd n_s^{(\phi)}}{\dd \ln k}
= -8 \left(\frac{2+p}{p^2}\right) \epsilon^2 = - 8 {\left( 2 + p \right) \over \left( 4N + p \right)^2 }
<0,
\end{align}
and
we also obtain the running in the case with the additional contribution from the curvaton fluctuation as
\footnote{One can show useful equations, $\frac{\dd \ln R}{\dd \ln k} = \frac{4}{p}\epsilon +\frac{2\mseff^2}{3H^2}$ and $-\frac{\dd}{\dd\ln k}\frac{1}{1+R}=\frac{\dd}{\dd\ln k}\frac{R}{1+R}=\frac{R^2}{(1+R)^2}\left[\frac{4}{p}\epsilon+\frac{2\mseff^2}{3H^2}\right]$.}
\begin{align}
n_s'(\ms)
&=-\frac{8}{1+R} \left[\frac{2+p}{p^2}\epsilon^2\right]
+\frac{R}{1+R}\left[-\frac{8}{p}\epsilon^2+\frac{4\ms^2}{3H^2}\epsilon\right]
+\frac{R}{(1+R)^2}\left[\frac{4}{p}\epsilon+\frac{2\ms^2}{3H^2}\right]^2,
\notag\\
n_s'(c_i)
&=-\frac{8}{1+R} \left[\frac{2+p}{p^2}\epsilon^2\right]
+\frac{R}{1+R}\left[-\frac{8}{p}\epsilon^2\right]
+\frac{R}{(1+R)^2}\left[\frac{4}{p}\epsilon+\frac{2}{3}c_i\right]^2,
\label{running2}
\end{align}
where $n_s'(\ms)$ denotes the model with the intrinsic mass, $\ms$, while
$n_s'(c_i)$ denotes the model with the HIM, $c_i H_*^2$, during inflation.
The first, the second and the third term represent
the contributions from the inflaton, the curvaton and their mixture, respectively.
Notice the second term has the extra term in the former model because in the latter model, $\mseff^2/H_*^2$ is constant and its
derivative vanishes. 

%
%
\begin{figure}[tbp]
  \hspace{-2mm}
  \includegraphics[width=75mm]{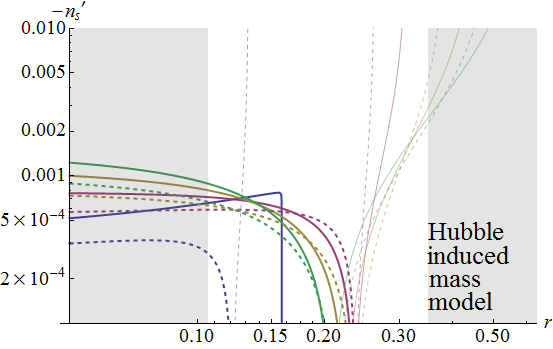}
    \hspace{5mm}
    \includegraphics[width=75mm]{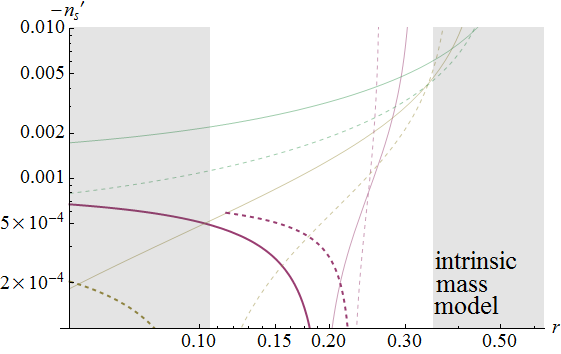}
 \caption
 {The running of the spectrum index $n_s'$ in the HIM model (left panel) and the intrinsic mass model (right panel). The vertical axis is $-n_s'$ while transparent lines represent positive $n_s'$.  The colors denotes the inflation model as
$p=2$ (blue), 4 (red), 6 (yellow) and 8 (green).
The e-folding number $N$ is 50 (solid lines) and 60 (dashed lines).
$n_s = 0.96$ is used to fix the curvaton (effective) mass.
The magnitude of the negative running is less than $10^{-3}$
while the positive running can be larger.
}
 \label{run plot}
\end{figure}
Fig.~\ref{run plot} shows $n_s'$ in the case with
the Hubble induced mass (left panel) and the intrinsic mass (right panel).
The mass of the curvaton, $\mseff^2$, would be determined by the observables with fixing the inflation model as shown in Eq. (\ref{mseff}),
 and hence from the expression (\ref{running2}), $n_s'$ can be also written in terms of the observables ($n_s, r, H_*$) and the inflationary parameter $(p, N)$.
As shown in Eq. (\ref{running2}), since terms originated from the mass of the curvaton $\sigma$ always have the positive contributions to $n_s'$, they somewhat cancel the negative contribution from the inflaton
which is proportional to $\epsilon^2$. Thus when $n_s'$ is negative, its magnitude is comparable
or smaller than the original prediction of the inflation
and it means that the maximum absolute value of the negative $n_s'$ is 
given by ${n_s'}^{(\phi)}$ listed as
\begin{align}
N=50:\qquad&
{n_s'}_{(p=2)}^{(\phi)} = -7.8\times10^{-4},&
\quad&
{n_s'}_{(p=4)}^{(\phi)} = -1.1\times10^{-3}
\\
\qquad&
{n_s'}_{(p=6)}^{(\phi)} = -1.5\times10^{-3},&
\quad&
{n_s'}_{(p=8)}^{(\phi)} = -1.8\times10^{-3},
\\
N=60:\qquad&
{n_s'}_{(p=2)}^{(\phi)} = -5.5\times10^{-4},&
\quad&
{n_s'}_{(p=4)}^{(\phi)} =-8.1\times10^{-3}
\\
\qquad&
{n_s'}_{(p=6)}^{(\phi)} = -1.1\times10^{-3},&
\quad&
{n_s'}_{(p=8)}^{(\phi)} = -1.3\times10^{-3}.
\end{align}
%

\subsection{Resolution of the degeneracy on the $n_s'$-$(n_T/r)$ plane}

%
%
As shown in the previous subsection, the ratio $n_T/r$ fixes the relative contribution of the curvaton
or the other additional degrees of freedom (see eq.~\eqref{intrinsic r}).
In other words, it implies how the observed $r$ deviates from
the original prediction of a inflation model, namely ``the consistency relation".
Therefore if  observations  fix both $r$ and $n_T$ with sufficient accuracy,
the most favored inflation model at least in the light of the tensor perturbation is determined without an uncertainty from curvaton-like models.
However, the curvaton model 
is not identified because
various curvaton models (e.g. the HIM model or the 
intrinsic mass model) can produce the same amount of $R$.

Then we need another observable to distinguish curvaton models.
The running of the spectrum index $n_s'$ can be useful for the purpose.
As we see in fig.~\ref{run plot}, $n_s'$ predicted by models
are different and can work as a discriminator.
This is because the parameters in curvaton models 
are fixed to reproduce the observed $n_s$. Thus the differences between models
appear in the prediction for $n_s'$. 

Let us consider the plane whose x-axis is $n_s'$ and y-axis is
$R = -1-8n_T/r$. 
Using eqs.~\eqref{r}, \eqref{tilt} and \eqref{running2},
the location on the $\{n_s', R\}$ plane can be described
by $n_s$ and $r$. In the two  curvaton models, 
the expressions of $n_s'$ are given by
\begin{align}
n_s' (\ms)
&=
2\epsilon(n_s-1+2\epsilon)+\frac{\epsilon}{2p}\left(r-16\epsilon\right)-\frac{r\epsilon}{p^2}
-\frac{r p^{-2}}{r-16\epsilon}\left[
4\epsilon+p(n_s-1+2\epsilon)
\right]^2,
\notag
\\
n_s'(c_i)
&=
\frac{p^2 r (n_s+2 \epsilon -1)^2+8 p \epsilon  \left(r (n_s+3 \epsilon -1)-16 \epsilon ^2\right)+r^2 \epsilon }{p^2 (16 \epsilon -r)},
\label{running4}
\end{align}
where $\epsilon = p / (4N + p)$ in the chaotic inflation model.
%
\begin{figure}[tbp]
  \hspace{-2mm}
  \includegraphics[width=75mm]{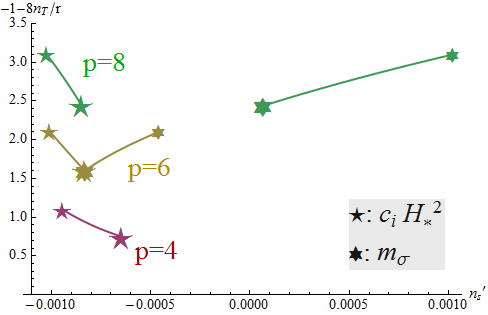}
    \hspace{3mm}
    \includegraphics[width=75mm]{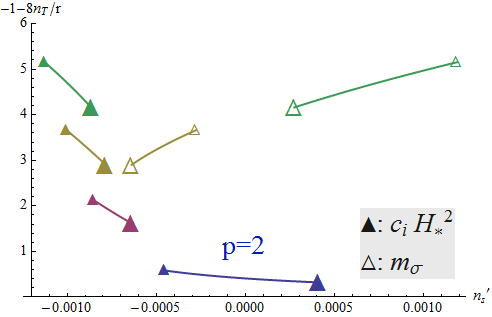}
 \caption
 {The left panel shows the model predictions for $\{n_s, r\}=\{0.945, 0.15\}$ (the red star in fig.~\ref{help}) while the right panel shows the case with
$\{n_s, r\}=\{0.95, 0.1\}$ (the red triangle in fig.~\ref{help}). The smaller symbols denote the $N=50$ cases
and bigger ones denote the $N=60$ cases.
One can see that model predictions are distinct from each other.
Only in the case of $p=6$, two models converge because the curvaton mass is negligible there (see fig.~\ref{help}).
}
 \label{stars}
\end{figure}
\begin{figure}[tbp]
  \hspace{-2mm}
  \includegraphics[width=75mm]{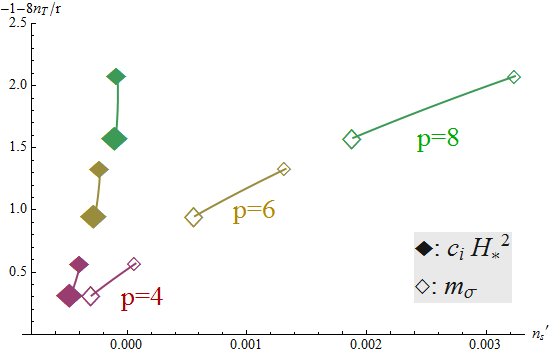}
    \hspace{3mm}
    \includegraphics[width=75mm]{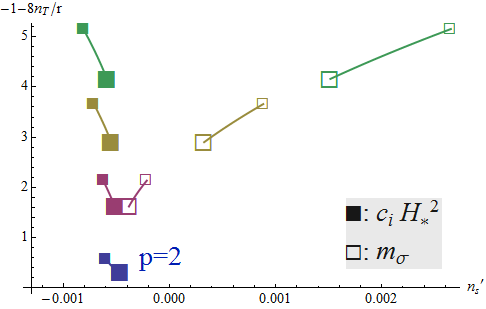}
 \caption
 {The left panel shows the case with $\{n_s, r\}=\{0.96, 0.2\}$ (the red diamond in fig.~\ref{help}) while the right panel shows the case with $\{n_s, r\}=\{0.965, 0.1\}$ (the red spade). 
}
 \label{spades}
\end{figure}

In fig.~\ref{stars} and fig.~\ref{spades}, we demonstrate that the two curvaton models in the $\phi^p$ chaotic inflation are distinguishable on the new plane.
These figures should be compared with fig.~\ref{help}.
The deviations between the original predictions of the inflation models
and the fiducial points (we adopt $\{n_s, r\}= \{0.945,0.15\}, \{0.95,0.1\}, \{0.96,0.2\}$ and $\{0.965,0.1\}$, see the red points in fig.~\ref{help}) reflect the differences on the $n_s'$-$R$ plane. It should be noted that the consistency relation
is satisfied on the x-axis of the $n_s'$-$R$ plane, namely $R=0$.

Regarding observations, in order to discriminate the curvaton models,
typically $\mathcal{O}(10^{-4})$ accuracy in the determination
of $n_s'$ is required. Such a great accuracy 
may be still challenging in the next generation observations~\cite{Dent:2012ne, Kohri:2013mxa}.
For example, in ref.~\cite{Kohri:2013mxa}, the authors argue that
1$\sigma$ uncertainty for $n_s'$ is larger than $10^{-4}$ even from
a combination of CMB and 21cm fluctuation future experiments (i.e. CMBpol + Omniscope). Thus a more advanced observation  is
required to distinguish the curvaton models.


\section{Comments on non-Gaussianity and the curvaton field value}
\label{NG}

In this section we discuss the non-linearity parameters and the curvaton field value $\sigma_*$.
In curvaton models with a quadratic potential,
the non-linearity parameters are given by~\cite{Ichikawa:2008iq, Enqvist:2013paa}
\begin{align}
\fnl &\simeq \ \frac{5}{6}\left( \frac{R}{1+R} \right)^2 
\frac{N_{\sigma\sigma}}{N_\sigma^2}
=\frac{5}{12}\left( \frac{R}{1+R} \right)^2\left[-3+\frac{4}{\hatr}+\frac{8}{4+3\hatr} \right],
\label{fnl}
\\
\tau_{\rm NL} &\simeq \left( \frac{R}{1+R} \right)^3
\left(\frac{N_{\sigma\sigma}}{N_\sigma^2} \right)^2
=\left( \frac{1+R}{R} \right) \left( \frac{6}{5}\fnl\right)^2,
\end{align}
where $\hatr$ is defined as the ratio between the radiation energy density $\rho_\gamma$
and the curvaton energy density $\rho_\sigma$ at the time of the curvaton decay:
\begin{equation}
\hatr \equiv \frac{\rho_\sigma}{\rho_\gamma}(t_{\sigma,{\rm dec}}),
\end{equation}
%
and the  contributions to the non-linearity parameters from the inflaton are ignored because they are suppressed by
the slow-roll parameters.
Note that $\fnl$ and $R$ should be considered together
because they are functions of $\hatr$.
In terms of $\hatr$, the parameter $R$ is given by~\cite{Lyth:2002my, Kawasaki:2011pd}
\begin{align}
R=\left(\frac{2 \hatr}{4+3\hatr}\right)^2 \frac{2\epsilon\Mpl^2}{\sigma_*^2}.
\label{rhat2}
\end{align}
Since $\hatr$ is positive and hence $2\hatr/(4+3\hatr)<2/3$,
an upper bound on $\sigma_*$ can be derived as,
\begin{equation}
\frac{\sigma_*}{\Mpl}< 
\frac{2}{3} \sqrt{\frac{2\epsilon}{R}},
\label{sigma upper limit}
\end{equation}
The reason why the upper limit on $\sigma_*$ is obtained is that a larger $\sigma_*$
leads to a smaller $\mcP_\zeta^{(\sigma)}$ which is inadequate even if 
the curvaton dominates the universe, $\hatr\to \infty$.
%
%
\begin{figure}[tbp]
  \hspace{-2mm}
  \includegraphics[width=75mm]{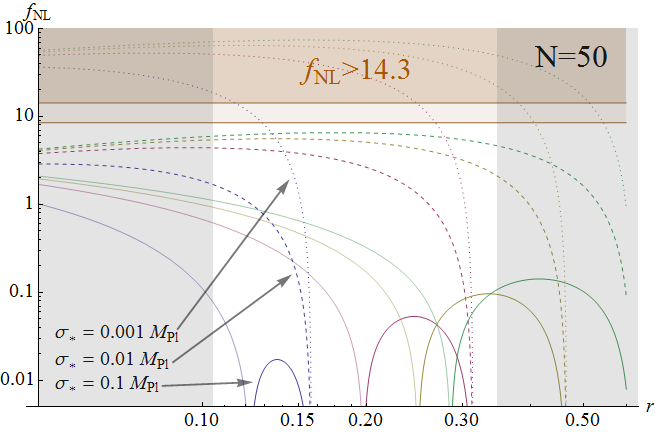}
    \hspace{3mm}
    \includegraphics[width=75mm]{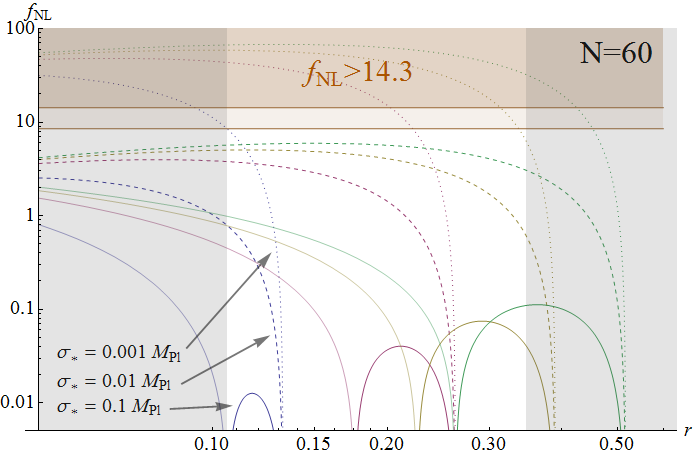}
 \caption
 {$\fnl$ for varying $\sigma_*$.
 The solid, dashed and dotted lines represent
 $\sigma_* = 10^{-1}, 10^{-2}$ and $10^{-3}\Mpl$, respectively.
 The transparent solid lines represent negative $\fnl$ values.
 The colors denotes the chaotic inflation model of
 $p=2$ (blue), 4 (red), 6 (yellow) and 8 (green).
 The e-folding number is $N=50$ (left panel) and $N=60$ (right panel).
 The brown (light brown) shaded regions are excluded by the Planck
 observation at the $2\sigma$ $(1\sigma)$ level.
 }
\label{fnl plot}
\end{figure}
\begin{figure}[!t]
  \hspace{-2mm}
  \includegraphics[width=75mm]{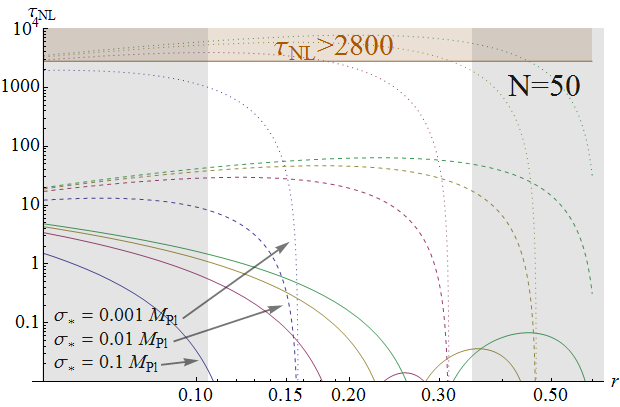}
    \hspace{3mm}
    \includegraphics[width=75mm]{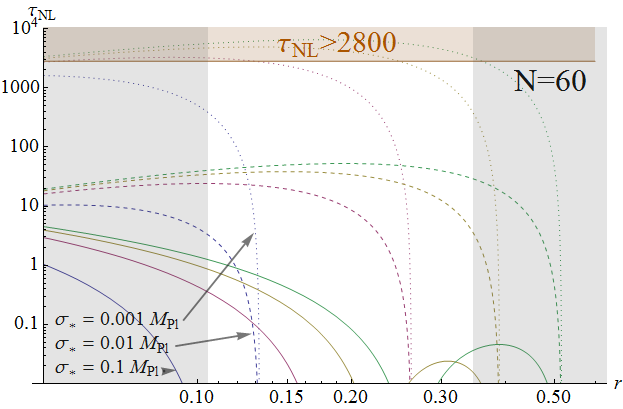}
 \caption
 {$\taunl$ for varying $\sigma_*$.
 All colors and plot styles have the same meaning as fig.~\ref{fnl plot}. 
}
 \label{taunl plot}
\end{figure}
%
By eliminating $\hatr$ from eqs.~\eqref{fnl} and \eqref{rhat2},
we obtain 
\begin{align}
\fnl&=\frac{5}{12}\left( \frac{R}{1+R} \right)^2
\left[-4+\frac{2\Mpl}{\sigma_*}\sqrt{\frac{2\epsilon}{R}}-\frac{3\sigma_*}{\Mpl}\sqrt{\frac{R}{2\epsilon}} \right],\label{sigma value}\\
\frac{\sigma_*}{\Mpl} &=
\sqrt{\frac{2\epsilon}{R}}\left[
\frac{R^{-2}}{15}\sqrt{36(1+R)^4\fnl^2+120R^2(1+R)^2\fnl+250R^4}\right.
\left.-\frac{2}{5}\left(\frac{1+R}{R}\right)^2\fnl-\frac{2}{3}\right].
\notag
\end{align}
%
This pair of equations are useful in two ways.
First, $\fnl$ (and hence $\taunl$) is given if $\sigma_*$ and a inflation model are fixed.
Second, since $\sigma_*$ decreases as $\fnl$ increases,
the lower bound on $\sigma_*$ can be obtained from the Planck constraint on $\fnl$~\cite{Ade:2013ydc}.
\footnote{
it is interesting to note that
by combining eqs.~\eqref{sigma upper limit} and \eqref{sigma value}, one can find the lower bound on $\fnl$ as
$\fnl > (5/12)(R/(1+R))^2
\left[ \sqrt{R/2\epsilon}-3\right].
$}

\begin{figure}[tbp]
\begin{center}
  \includegraphics[width=100mm]{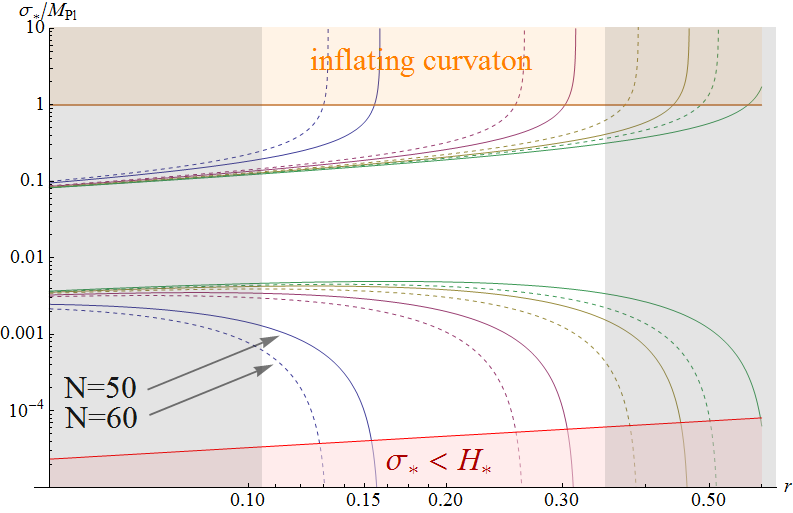}
\end{center}
 \caption
 {The regions between the pairs of the same colored lines are allowed for $\sigma_*$. The upper limits is coming from eq.~\eqref{sigma upper limit} and the lower limit is originated from eq.~\eqref{sigma value} with $\fnl>14.3$. The constraints are relaxed where $R$ becomes tiny. However the conditions for the  validity of the perturbation $\sigma_*>H_*$
and the  non-inflating curvaton $\sigma_*<\Mpl$ do not allow an arbitrary value of $\sigma_*$ for any $r$.}
 \label{sigma bound}
\end{figure}

In fig.~\ref{fnl plot} and fig.~\ref{taunl plot},
we plot $\fnl$ and $\taunl$ for varying $\sigma_*$. 
In the chaotic inflation, $\sigma_*$ is naturally expected to be lower than $\Mpl$.
This is because in the initial condition of the chaotic inflation,
a light scalar field can take a large value that is even more than $\Mpl$~\cite{Linde:1983gd,Linde:2005ht},
but if $\sigma_* \gtrsim \Mpl$, $\sigma$ becomes not a curvaton but an inflaton. 
Therefore the cases with $\sigma_*=10^{-1}\Mpl, 10^{-2}\Mpl$ and $10^{-3}\Mpl$
are shown in those figures.
Basically, a larger non-linearity are produced by a smaller $\sigma_*$.
However, due to the Planck constraint $\fnl<14.3$ and $\taunl<2800$ at the $2\sigma$ level, 
$\sigma_*$ which is less than $10^{-3}\Mpl$ is excluded expect for the narrow
range of $r$. This result is consistent with fig.~\ref{sigma bound}.

Combining eqs.~\eqref{sigma upper limit} and \eqref{sigma value} with $\fnl<14.3$,
we obtain the allowed value of $\sigma_*$.
That is plotted in fig.~\ref{sigma bound}.
In the figure, one can see that $\sigma_*$ is typically restricted into one or two orders of magnitude.
Nevertheless the allowed value around $10^{-2}\Mpl$
does not need a extreme fine-tuning in the context of the chaotic inflation.

\section{Conclusion}
\label{Conclusion}

In this paper, we comprehensively investigate the curvaton model with
the quadratic potential in the chaotic inflation models 
whose potentials are $\phi^p$ with
$p=2,4,6$ and 8. With the observation
of the tensor-to-scalar ratio $r$ by BICEP2 experiment, the contribution of
a curvaton-like component to the scalar perturbation can be calculated
in the model-independent manner (fig.~\ref{R plot}).
Specifically we consider the two models of curvaton,
that the Hubble induced mass curvaton model and the intrinsic mass curvaton model. Then the curvaton mass is also computed (fig.~\ref{ci plot}).

Although the two parameter space of $n_s$ and $r$ is useful to discriminate
inflation models, we show the two curvaton models in the chaotic inflation models are heavily degenerated on the plane (fig.~\ref{help}).
However, it is found that the running of the spectrum index $n_s'$
are different between the two models (fig.~\ref{run plot}). This is because the model parameter
(i.e. the curvaton mass) is fixed by the spectrum index.
Moreover, the inflation models can be well determined by the observation
of the tensor tilt $n_T$. Since $n_T$ is given only by the inflation dynamics,
the ratio of $n_T$ and $r$, or "the consistency relation", 
implies the amount of the additional contribution from the curvaton.
Therefore, introducing the new 2 parameter space of $n_s'$ and $n_T/r$,
we demonstrate the degeneracy of the two curvaton models can be resolved
(fig.~\ref{stars}, \ref{spades}).
To distinguish the curvaton models, a precise measurement of $n_s'$
with an $\mathcal{O}(10^{-4})$  uncertainty is required.

We also calculate the non-linearity parameters in the light of BICEP2 result
(fig.~\ref{fnl plot}, \ref{taunl plot}).
Since $\fnl$ increases as the curvaton field value during inflation $\sigma_*$
decreases and Planck satellite put the upper bound on it, $\fnl< 14.3$, 
the upper bound on $\sigma_*$ is obtained. On the other hand, $\delta N$
induced by the curvaton modulation decreases as $\sigma_*$ increases and
the curvaton contribution is fixed by $r$, the lower bound on $\sigma_*$ is also obtained. We found that $\sigma_*$ is constrained into a narrow range, but its value $\sim 10^{-2}\Mpl$ is natural in the chaotic inflation (fig.~\ref{sigma bound}).


\acknowledgments

We would like to thank Ryusuke Jinno for his useful comments.
This work is supported by Grant-in-Aid for Scientific research from the Ministry of Education, Science, Sports, and Culture (MEXT), Japan, No. 25400248 (M.K.), No. 21111006 (M.K.) and the World Premier International
Research Center Initiative (WPI Initiative), MEXT, Japan.
T.F. and S.Y. acknowledge the support by Grant-in-Aid for JSPS Fellows
No. 248160 (TF) and No. 242775 (SY).

\appendix
\section{The other parameters}
\label{The other parameters}

In this appendix, we discuss the remaining model parameters,
namely $\hatr, \sigma_*$ and $\ms/\Gamma_\sigma$.
They are fixed if $\fnl$ and $r$ as well as $p$ and $N$ are given.

\begin{figure}[tbp]
  \hspace{-2mm}
  \includegraphics[width=75mm]{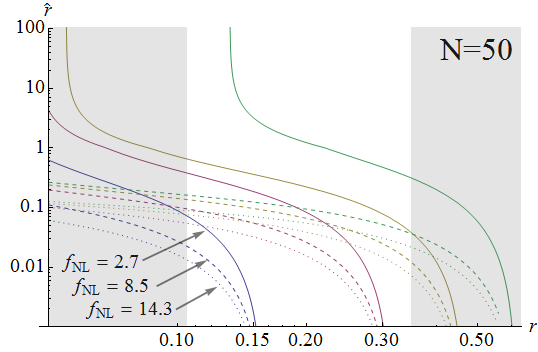}
    \hspace{5mm}
    \includegraphics[width=75mm]{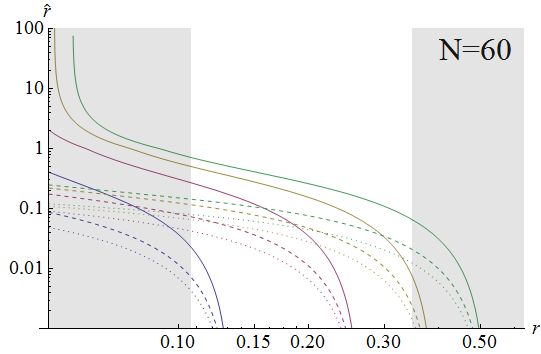}
 \caption
 { The plot of $\hatr \equiv \rho_\sigma/\rho_\gamma(t_{\sigma,{\rm dec}})$.
 The colors denotes the chaotic inflation model of
$p=2$ (blue), 4 (red), 6 (yellow) and 8 (green).
The e-folding number is $N=50$ (left panel) and $N=60$ (right panel). Contrary to the pure curvaton case, $\hatr$ can be smaller than 0.1 when the curvaton contribution is less than the inflaton's ($R\lesssim1$).
}
 \label{rhat plot}
\end{figure}

First let us calculate $\hatr$. By solving eq.~\eqref{fnl} with respect to $\hatr$, 
we can obtain the expression of $\hatr$ as
\begin{equation}
\hatr =\frac{4}{9}\frac{ \sqrt{36 \fnl^2 (R+1)^4+120 \fnl (R+1)^2 R^2+250 R^4}-6 \fnl (R+1)^2+5 R^2}
{4 \fnl (R+1)^2+5 R^2}.
\end{equation}
In fig.~\ref{rhat plot}, $\hatr$ is shown for $\fnl=2.7, 8.5$ and $14.3$. 
In the pure curvaton case ($R\to \infty$ in eq.~\eqref{fnl}),
one obtains the lower bound on $\hatr$ as
\begin{equation}
\fnl=\frac{5}{12}\left[-3+\frac{4}{\hatr}+\frac{8}{4+3\hatr} \right]< 14.3
\quad \Longrightarrow \quad
\hatr > 0.11,
\qquad(R\gg1).
\end{equation}
However, in the mixed case, since the $R$ factor in eq.~\eqref{fnl} changes the $\hatr$ dependence, a small $\hatr\ (<0.11)$ is allowed and the curvaton does not have to almost dominate the universe.

\begin{figure}[tbp]
  \hspace{-2mm}
  \includegraphics[width=75mm]{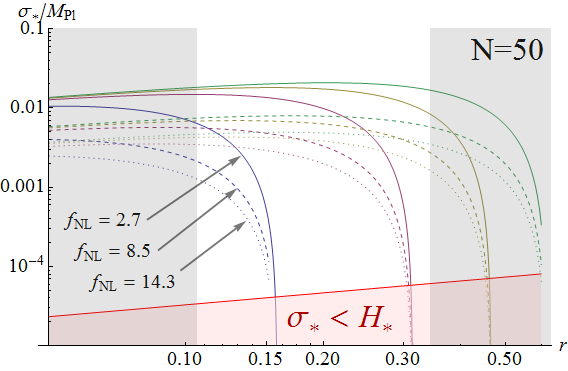}
    \hspace{5mm}
    \includegraphics[width=75mm]{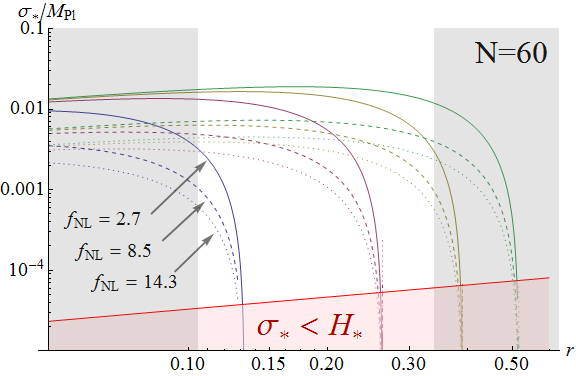}
 \caption
 { The $\sigma_*$ value computed from the observables (i.e. $\mcP_\zeta, r$ and $\fnl$) and the inflation model (i.e. $\phi^p$ chaotic inflation for $n=2,4,6$ and $8$ from bottom to top). The e-folding number is
$N=50$ (left panel) and $N=60$ (right panel). The solid, dashed, dotted lines represent
$\fnl = 2.7, 8.5, 14.3$, respectively. In the red shaded region, 
the perturbation of the curvaton is not reliable.
The dotted lines ($\fnl=14.3$) can be understood as the lower bound on
$\sigma_*$.
$\sigma_*$ is typically smaller than the Planck scale by a few orders of magnitude.
.
}
 \label{sigma plot}
\end{figure}

Second, let us see the value of $\sigma_*$. It is given in eq.~\eqref{sigma value} as the function of $\fnl$.
In fig.~\ref{sigma plot}, $\sigma_*$ is plotted for varying $\fnl$.
It is a counterpart of fig.~\ref{fnl plot}.

\begin{figure}[tbp]
  \hspace{-2mm}
  \includegraphics[width=75mm]{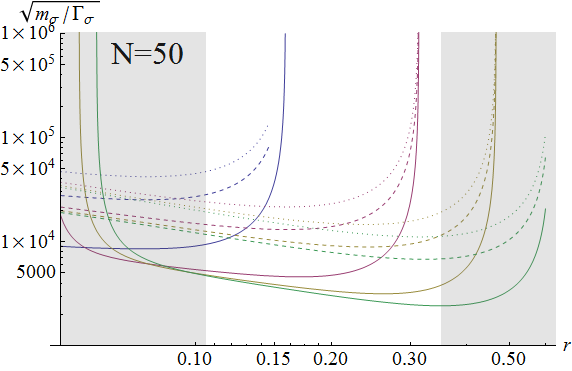}
    \hspace{3mm}
    \includegraphics[width=75mm]{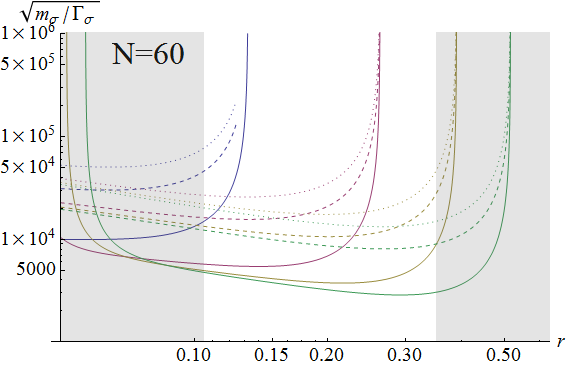}
 \caption
 { The square root ratio between the curvaton intrinsic mass $\ms$ and its decay rate $\Gamma_\sigma$. The solid, dashed, dotted lines represent
$\fnl = 2.7, 8.5, 14.3$, respectively. The colors denotes the chaotic inflation model of
$p=2$ (blue), 4 (red), 6 (yellow) and 8 (green).
$\ms<\Gamma_\phi$ is assumed.
As $\fnl$ increases,  $\sqrt{\ms/\Gamma_\sigma}$
also increases because $\sqrt{\ms/\Gamma_\sigma}$ is inversely proportional to $\hatr<\mathcal{O}(1)$.
}
 \label{mu}
\end{figure}

Finally, the ratio between the curvaton intrinsic mass and its decay rate is also written as a function of $\fnl$.
It is known that $\hatr$ is given by~\cite{Kawasaki:2011pd}
\footnote{If the curvaton starts to oscillate before the inflaton decays,
an extra factor $\sqrt{\Gamma_\phi/\ms}$ is multiplied. 
Moreover if $\hatr >1$,
the expression should be raised to the power of 3/4.}
\begin{equation}
\hatr 
\simeq \frac{V(\sigma_{\rm osc})}{3\Mpl^2 H_{\rm osc}^2}
\sqrt{\frac{m_\sigma}{\Gamma_\sigma}}
\simeq \frac{\sigma_*^2}{6\Mpl^2}\sqrt{\frac{m_\sigma}{\Gamma_\sigma}},
\qquad
(\hatr<1,\ \ms<\Gamma_\phi),
\label{rhat appendix}
\end{equation}
where the subscript ``osc" denotes the onset of the curvaton oscillation, and $\Gamma_\sigma$ and $\Gamma_\phi$ are the decay rates of the curvaton
and the inflaton, respectively.
Here we approximate $\sigma_{\rm osc} \simeq \sigma_*$ by neglecting the curvaton evolution until the onset of its oscillation.
Substituting eq.~\eqref{rhat appendix} into eq.~\eqref{rhat2}, we obtain 
\begin{equation}
\sqrt{\frac{m_\sigma}{\Gamma_\sigma}}= \frac{\Mpl}{\sigma_*}\frac{24}{2\sqrt{2\epsilon/R}-3\sigma_*/\Mpl},
\end{equation}
and since $\sigma_*$ is given in eq.~\eqref{sigma value}, 
$\sqrt{\ms/\Gamma_\sigma}$ can be also explicitly given as a function of $\epsilon, R$ and $\fnl$.
While we omit  its complicated expression, fig.~\ref{mu}
shows its value.
In fig.~\ref{mu}, one can see that an increasing $\fnl$ leads to a larger
$\sqrt{\ms/\Gamma_\sigma}$. Apparently, however, this behavior is
against the instincts because one may expect that a larger $\fnl$
means a smaller $\hatr$ and the requirement to boost $\hatr$
by the hierarchy between $\ms$ and $\Gamma_\sigma$ is relaxed.
Nevertheless, note that eq.~\eqref{rhat2} fixes the relation between
$\hatr$ and $\sigma_*$ as $\hatr \propto \sigma_*$ for $\hatr<\mathcal{O}(1)$.
Then eq.~\eqref{rhat appendix} implies $\sqrt{\ms/\Gamma_\sigma} \propto \hatr^{-1}$.
Therefore although it seems $\hatr \propto \sqrt{\ms/\Gamma_\sigma}$
in eq.~\eqref{rhat appendix}, the actual dependence is in inverse proportion.   

In the intrinsic mass curvaton model, $\mseff = \ms$, we have already computed the $\ms$ value 
in eq.~\eqref{mseff} and fig.~\ref{ci plot}, and we can obtain $\Gamma_\sigma$ explicitly as
\begin{equation}
\Gamma_\sigma = \frac{\ms\sigma_*^2}{576\Mpl^2} 
\left(2\sqrt{\frac{2\epsilon}{R}}-\frac{3\sigma_*}{\Mpl}\right)^2.
\end{equation}
%

%

\end{document}